\documentclass[aps,prb,twocolumn,reprint,superscriptaddress,showkeys,floatfix,amsmath,amssymb]{revtex4-1}
\usepackage{float}

\usepackage[utf8]{inputenc}
\usepackage{mathrsfs}
\usepackage{color}
\usepackage{hyperref}
\usepackage{graphicx}
\usepackage{xcolor, soul}
\usepackage{dcolumn}
\usepackage{bm}
\usepackage{natbib}

\begin{document}
	
	\preprint{APS/123-QED}
	
	\title[]{Effect of applied pressure on the non-relativistic spin-splitting (NRSS) of \texorpdfstring{FeSb$_{2}$}{FeSb2} altermagnet: A first-principles study }
	

	\author{Shalika R. Bhandari}
	\email [Shalika Ram Bhandari:] {shalikram.bhandari@bmc.tu.edu.np}
	\affiliation{Department of Physics, Bhairahawa Multiple Campus, Tribhuvan University, Siddarthanagar-32900, Rupandehi, Nepal}
	
	\author{R. Tamang}
	\affiliation{Department of Physics, Mizoram  University, Aizawl 796004, India}
	
	\author{Keshav Shrestha}
	\email [K Shrestha:] {kShrestha@wtamu.edu}	
	\affiliation{Department of Chemistry and Physics, West Texas A\&M University, Canyon, Texas 79016, USA}
	
	\author{Samy Brahimi}
	\affiliation{Laboratoire de Physique et Chimie Quantique, Universite Mouloud Mammeri de Tizi-Ouzou, 15000 Tizi-Ouzou, Algeria}
	\affiliation{Peter Gr\"unberg Institute, Forschungszentrum J\"ulich and JARA, J\"ulich, Germany}
	\author{Samir Lounis}
	\affiliation{Institute of Physics, Martin-Luther-University Halle-Wittenberg, 06099 Halle (Saale), Germany}
	
	\author{D. P. Rai}
	\email [D P Rai:] {dibyaprakashrai@gmail.com}
	\affiliation{Department of Physics, Mizoram  University, Aizawl 796004, India}
	\affiliation{Peter Gr\"unberg Institute, Forschungszentrum J\"ulich and JARA, J\"ulich, Germany}
	%
	
	

	\date{\today}
	
	\begin{abstract}

		We have investigated the pressure-dependent electronic structure, phonon stability, and anomalous Hall response of the recently discovered altermagnet FeSb$_2$ from density functional theory (DFT) and Wannier function analysis. From density functional perturbation theory (DFPT) calculations, we have found that FeSb$_{2}$ remains dynamically stable up to 10 GPa, evidenced by positive phonon frequencies. Our spin-polarised band structure shows that the node of band crossing between spin-up and spin-down bands around the Fermi energy exactly lies at the $\Gamma$ and A-symmetry points. The Fermi crossing is mostly exhibited by band-24, band-25 and band-26. The non-relativistic spin-splitting (NRSS) along M'-$\Gamma$-M and A-Z-A' symmetry is attributed to the broken time-reversal ($\mathcal{PT}$) symmetry. There are significant changes in the band profile under applied pressure, as one can see the shifting of the node of band-24 and band-26 towards the lower energy side. The NRSS exhibited by band-24 along M'-$\Gamma$-M symmetry is notably small. 
		Although the strength of NRSS of band-26 along A-Z-A' symmetry is significant but reduces under applied pressure. The anomalous Hall conductivity (AHC) values are prominent in -1 to 1 eV range. A sharp peaked and positive AHC values at ambient pressure, becomes spectrally broadened and negative at 10 GPa due to pressure-induced band crossings and redistribution of Berry curvature near the Fermi level. We have observed that the values of spin hall conductivity (SHC) are around 2-2.5 times lower as compared to AHC and prominent in between -1.0 eV to 1.0 eV. Our results establish FeSb$_{2}$ as a tunable altermagnetic candidate where pressure can modulate both topological transport and dynamic stability, offering opportunities for strain-engineered Hall responses in compensated magnetic systems.
	\end{abstract}
	
	\maketitle
	

	\section{Introduction}
	The recent discovery of altermagnetism has garnered significant interest in condensed matter physics due to its potential in spintronics, particularly magnetic sensing and scalable data storage \cite{vsmejkal2022beyond,vsmejkal2022emerging,song2025altermagnets}. Altermagnets  (AMs) are a new class of collinear magnetic materials that exhibit spin-split electronic bands despite having zero net magnetization. This symmetry-protected spin splitting arises from the combined breaking of time-reversal ($\mathcal{T}$) and inversion symmetries, making AMs promising for dissipationless spintronic applications \cite{vsmejkal2022emerging,vsmejkal2022beyond,song2025altermagnets,tamang2025altermagnetism,vsmejkal2022giant}. Uniquely, this spin splitting does not rely on spin-orbit coupling (SOC), but instead stems from motif-pair anisotropy, enabling unconventional spin polarization \cite{vsmejkal2020crystal,hayami2019momentum,yuan2023degeneracy,sattigeri2023altermagnetic,mazin2023altermagnetism,vsmejkal2023chiral,vsmejkal2022beyond,mcclarty2024landau,mazin2022altermagnetism,gomonay2018antiferromagnetic,shao2021spin,gonzalez2023spontaneous,mazin2023induced}. Altermagnets thus bridge the gap between ferromagnets (FM) and antiferromagnets (AFM), offering the spin functionality of the former and the stability of the latter. Recently identified AMs materials with collinear AFM order and nonrelativistic spin splittings (NRSS) from the experiments include RuO${_2}$ \cite{fedchenko2024observation,lin2024observation}, MnTe\cite{lee2024broken,krempasky2024altermagnetic,osumi2024observation}, CrSb\cite{ding2024large,reimers2024direct,yang2025three,zeng2024observation}, KV$_2$Se$_2$O\cite{jiang2023prediction}, CoNb$_4$Se$_8$\cite{dale2024non,sakhya2025electronic}, and others.\\
	
	Recently, we conducted first-principles electronic structure studies of FeSb$_{2}$\cite{phillips2025electronic}, revealing distinct asymmetry between spin-up and spin-down states in the spin-polarized electronic bands. Notably, the presence of non-relativistic spin splitting (NRSS) in this collinear AFM provides clear evidence of AMs in FeSb$_2$. Three bands cross the Fermi level (E$_F$), forming the Fermi surface of FeSb$_2$, and the band-resolved Fermi pockets exhibit $d$-wave symmetry with evident spin asymmetry. Furthermore, the quantum oscillation frequencies derived from the extremal cross-sectional areas of the Fermi surface are in good agreement with experimental de Haas-van Alphen (dHvA) frequencies.
	
	While the intrinsic magnetic and topological properties of AMs have recently attracted significant attention, their response to external stimuli such as pressure or strain remains relatively underexplored. Pressure engineering, a well-established technique in condensed matter physics, offers a clean and effective way to tune interatomic distances, lattice symmetry, and crystal-field environments without introducing chemical disorder \cite{chakraborty2024strain,fan2025high,devaraj2024interplay}. This method has proven to be a powerful tool for modulating electronic, magnetic, and topological properties across a wide range of quantum materials. For instance, Devaraj et al. employed first-principles calculations and spin-group analysis to demonstrate tunable altermagnetism in $\alpha$-MnTe and $\gamma$-MnTe under pressure, revealing enhanced spin-splitting and anomalous Hall conductivity (AHC) values exceeding $370~\mathrm{S/cm}$ and $400~\mathrm{S/cm}$, respectively\cite{devaraj2024interplay}. Similarly, Fan et al. reported stable altermagnetic ordering in MnF${_2}$ across a pressure range of $0$-20 GPa, with spin-splitting tunable up to 307.5 meV~\cite{fan2025high}. In FeSe, applied pressure enhances superconductivity from 8 K to over 35 K by modifying the nematic order and reconstructing the Fermi surface~\cite{medvedev2009electronic}. In Mn${_3}$Sn, uniaxial strain has been shown to control both the magnitude and sign of the AHC by reshaping the Berry curvature distribution~\cite{nayak2016large}. Likewise, the Dirac semimetal Cd${_3}$As${_2}$ undergoes a topological phase transition under pressure, driven by Dirac node annihilation~\cite{zhang2017room}.
	
	This study investigates the impact of external pressure on the electronic and phononic properties of FeSb$_2$ using first-principles calculations. Our phonon dispersion analysis reveals that FeSb$_2$ maintains positive phonon modes up to 10 GPa, confirming the system's dynamical stability within this pressure range. Additionally, electronic band structure calculations show a gradual suppression of spin-splitting near the Fermi level as pressure increases up to 10 GPa. We also examine how pressure influences the evolution of the Fermi surface, anomalous Hall conductivity (AHC), and spin Hall conductivity (SHC). 
	
	\section{
		Computational Details}
	
	For the spin-polarized electronic structure calculation, we have used an open-sourced DFT-based package called QUANTUM ESPRESSO (QE)\cite{giannozzi2009quantum,giannozzi2017advanced,giannozzi2020quantum}. A scalar non-relativistic pseudopotential within the projector-augmented wave (PAW) with Perdew-Burke-Ernzerhof (PBE) formalism of the generalized gradient approximation (GGA) is employed for all electrons exchange-correlation\cite{perdew1996generalized,perdew2007generalized}.
	A kinetic energy cutoff of 70 Ry and electron-density cutoff of 750 Ry is used along with the 'MV' smearing for the confinement of the plane waves, self-consistent (SC) and non-self-consistent (NSC) calculations. The threshold self-consistent field (SCF) convergence criteria is set to be 10$^{-6}$ Ry. For the integration of the first Brillouin zone, we have used the k-mesh of 12$\times$12$\times$12 within a Monkhorst-Pack scheme \cite{monkhorst1976special}, for the incorporation of strong correlated effects of Fe-$3d$ electrons, a Hubbard potential (U=2.0 eV) is included along with GGA as GGA+U. 
	To verify the structural and thermodynamic stability, we have calculated the phonon dispersion by using the Phonon software of master code QE that relies on Density Functional Perturbation Theory (DFPT)\cite{baroni2001phonons}. The phonon input was set by constructing a 2$\times $2$\times $2 grid of q-points in the Brillouin zone. Following the previous experimental report\cite{phillips2025electronic} an antiferromagnetic (AFM) coupling was set between Fe1 and Fe2 atoms of FeSb$_2$ [see Fig.\ref{fig:spin}], and structural optimization was performed with the option 'vc-relax' within the BFGS algorithm \cite{Broyden1970, Fletcher1970, Goldfarb1970, Shanon1970}. FeSb$_2$  is found to be stable with the lowest energy in the antiferromagnetic state, consistent with the experimental report\cite{mazin2021prediction}. 
	Furthermore, along with the electronic structure calculation, we have extended our calculation to explore the magneto-transport properties of the solid crystalline materials. The computational code Wannier90 is used for the calculation of anomalous and spin Hall conductivity \cite{MOSTOFI20142309}. By fitting the electronic band structure from the full-relativistic DFT calculation, Wannier90 will develop a tight-binding Hamiltonian and maximally localized Wannier functions (MLWFs). In the process, the delocalized Bloch wavefunctions from DFT are transformed to a localized basis set. This helps to explore the untraceable information about the electronic properties from DFT  by interpolating the bands across the Brillouin zone. From the interpolated Wannier bands, a geometrical phase of the wavefunctions can be extracted, which gives the Berry curvature \cite{nagaosa2010anomalous, Jungwirth_berry}. The AHC ($\sigma_{xy}^{AHC}$) and SHC ($\sigma^{SHC}$) for all occupied bands were evaluated by integrating a Berry curvature in the Brillouin zone over a dense k-point mesh using Kubo’s formula within linear response theory\cite{nagaosa2010anomalous, Yimin2022, Sinova2015}.

	\begin{equation}
		\Omega_{n}^{z}(\mathbf{k}) = -2, \text{Im} \sum_{m \neq n} \frac{ \langle u_{n\mathbf{k}} | \hat{v}x | u{m\mathbf{k}} \rangle \langle u_{m\mathbf{k}} | \hat{v}y | u{n\mathbf{k}} \rangle }{(E_{m\mathbf{k}} - E_{n\mathbf{k}})^2}  
	\end{equation}
	
	where $\Omega_{n}^{z}(\mathbf{k}$ is the Berry curvature for band ($n$),
	$u_{n\mathbf{k}}$ are the cell-periodic parts of Bloch functions, $\hat{v}_x$ and $\hat{v}_y$ are velocity operators, $E_{n\mathbf{k}}$ are the band energies.
	
	\begin{equation}
		\sigma_{xy}^{AHC} = -\frac{e^2}{\hbar} \int_{\text{BZ}} \frac{d\mathbf{k}}{(2\pi)^3} \sum_{n}^{\text{occ}} \Omega_{n}^{z}(\mathbf{k})\;,
	\end{equation}
	
	where ($e$) is the electron charge, $\hbar$ is the reduced Planck's constant, the sum runs over occupied bands.
	
	\begin{equation}
		\sigma^{SHC} = \frac{\hbar}{2e} \times \int_{\text{BZ}} \frac{d\mathbf{k}}{(2\pi)^3} \sum_{n}^{\text{occ}} \Omega_{n}^{s,z}(\mathbf{k})\;,
	\end{equation}
	
	where $\Omega_{n}^{s,z}(\mathbf{k})$ is the spin Berry curvature, computed similarly to the Berry curvature but incorporating spin operators.
	
	To calculate the anomalous Hall conductivity (AHC) and spin Hall conductivity (SHC), wannierization of the bands fitting was performed using a dense k-grid of 150$\times$150$\times$150, and the magnetization is considered along the (001) direction.

	\section{RESULTS AND DISCUSSION} 
	\subsection 
	{Crystal Structure and Phonon Analysis}
	
	\begin{figure}
		\centering
		\includegraphics[width=1.00\linewidth]{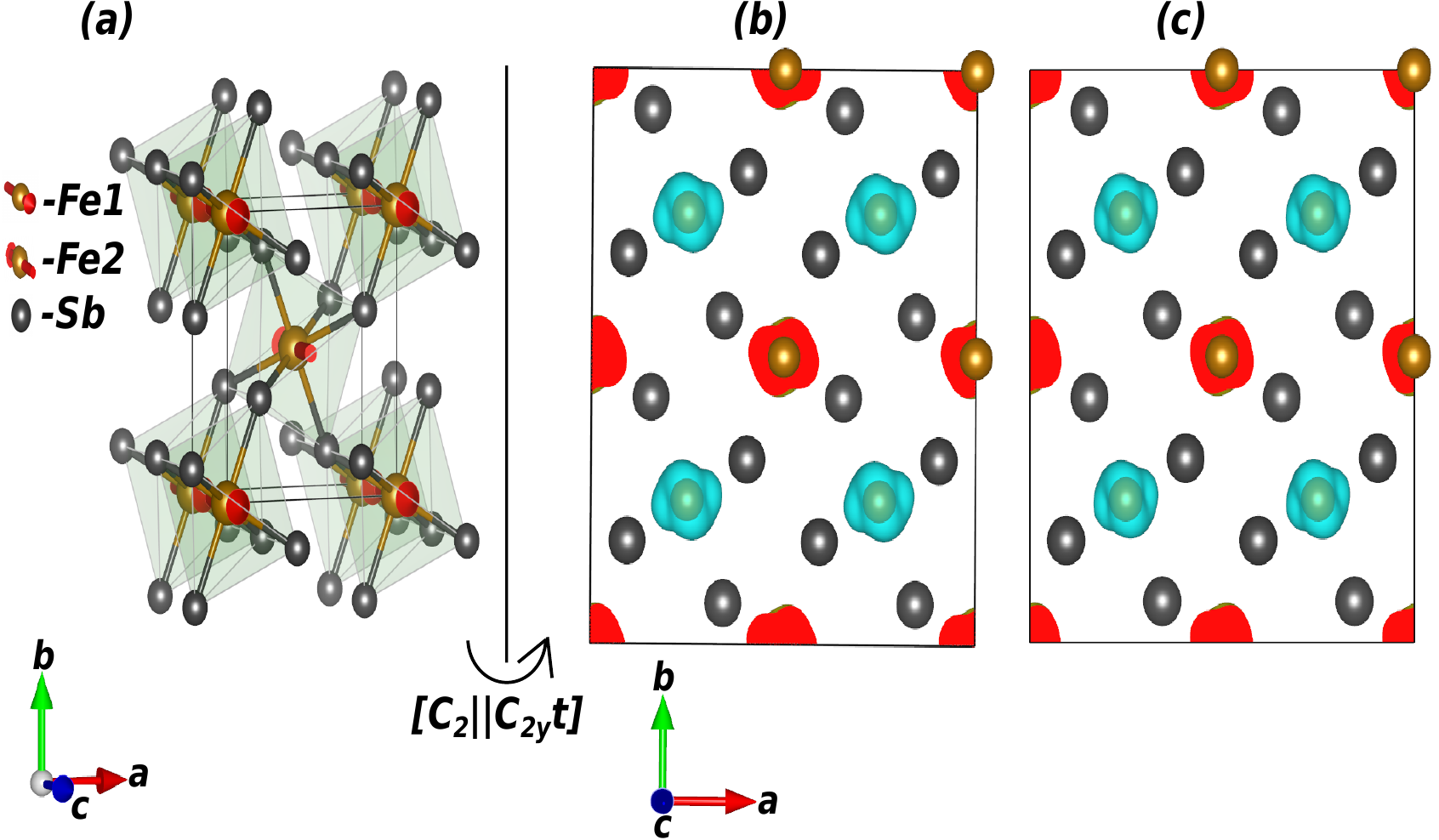}
		\caption{(a) Unit cell structure of FeSb$_2$ with anisotropic local environment and Spin-density at (b) 0.25 GPa,  (c) 10.0 GPa with isosurface value 0.0035 a.u. (Red colour $\uparrow$ and $\downarrow$ are the magnetic moments associated with Fe1 and Fe2 atoms and are distinguished by translucent orange and blue spin densities, respectively)}
		\label{fig:spin}  
	\end{figure}
	
	\begin{figure}[!ht]
		\centering
		\includegraphics[width=1.0\linewidth]{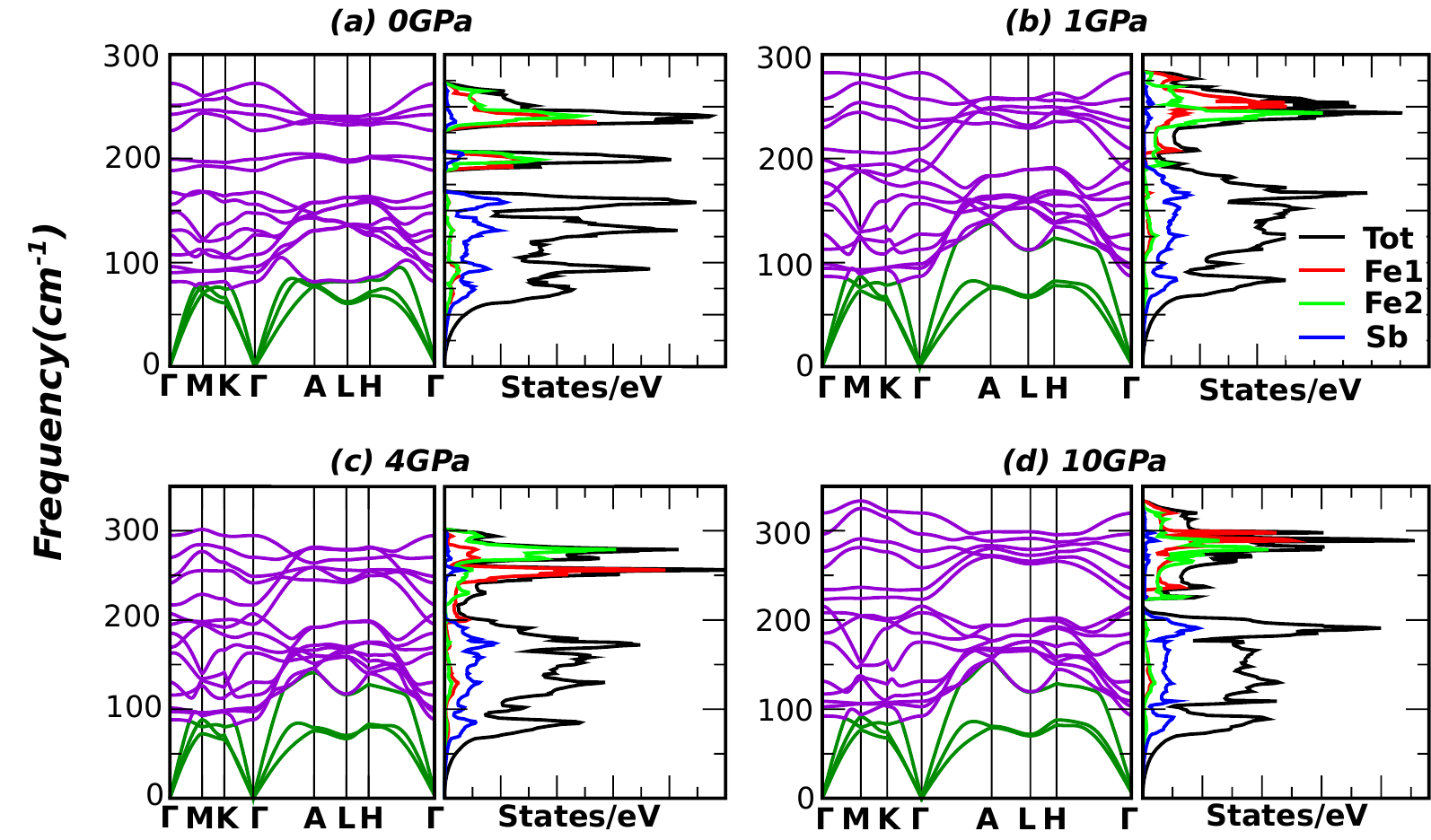} 
		\caption{(Color online) Phonon dispersion relations and phonon density of states (DOS) for FeSb${_2}$ under hydrostatic pressures of (a) 0.0 GPa, (b) 1.0 GPa, (c) 4.0 GPa, and (d) 10.0 GPa. The absence of imaginary frequencies in all panels confirms the dynamical stability of FeSb${_2}$ across this pressure range.} 
	\label{fig:phon}
\end{figure}

\begin{figure*}[htbp]
	\centering
	\includegraphics[width=1\linewidth]{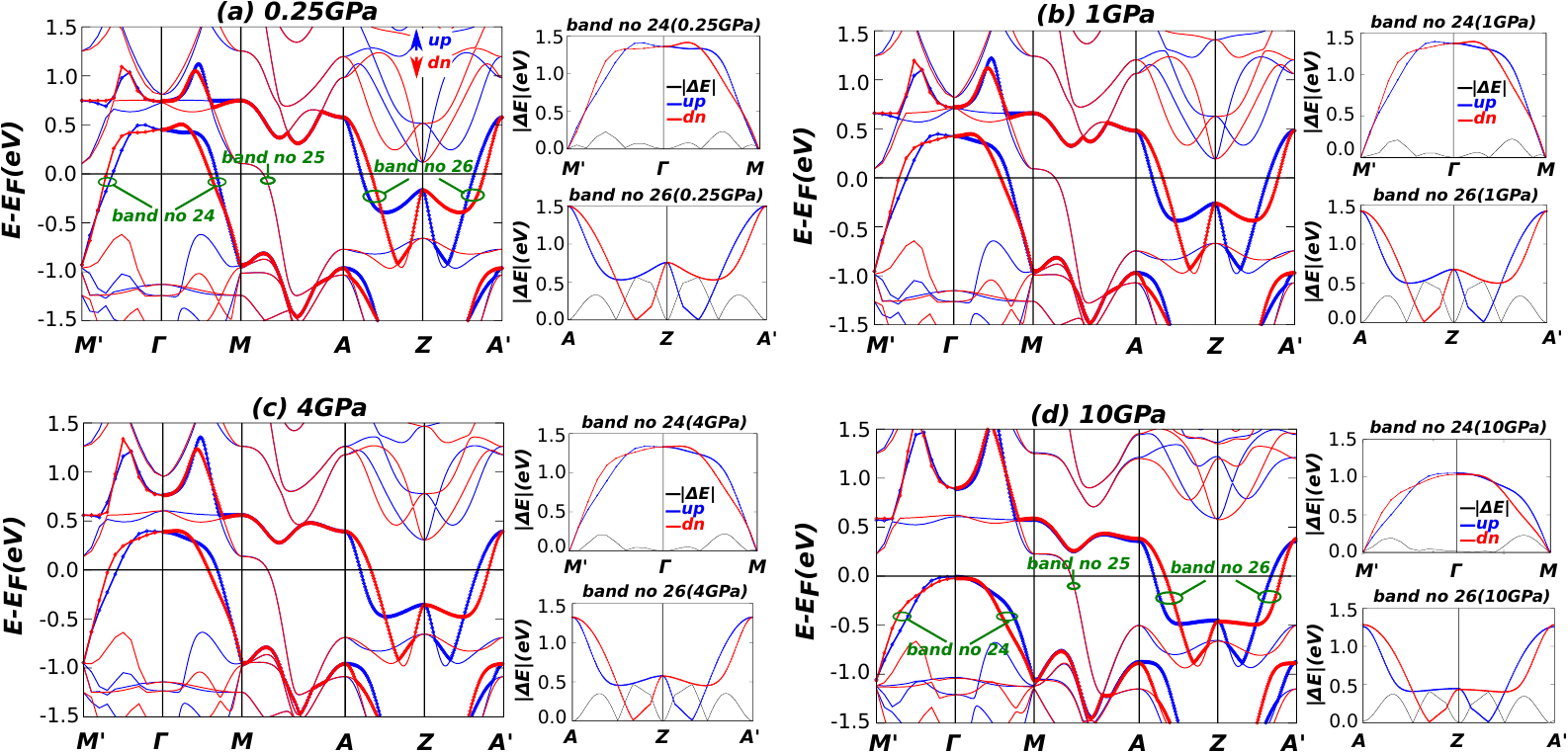}
	\caption{ Spin-resolved electronic band structures of FeSb${_2}$ under varying hydrostatic pressure: (a) 0.25 GPa, (b) 1.0 GPa, (c) 4.0 GPa, and (d) 10.0 GPa. red and blue lines represent spin-up and spin-down states, respectively. Spin-splitting in the absence of net magnetization indicates the presence of altermagnetic order.  The evolution of band dispersion and spin splitting with increasing pressure highlights the pressure dependence of altermagnetism in FeSb${_2}$. The Fermi level is set to zero. }
	\label{fig:band}  
\end{figure*}

The crystal structure of FeSb${_2}$, having space group P${\it nnm}$(58) with two antiferromagnetically coupled Fe1 and Fe2 atoms, is shown in  Fig.\ref{fig:spin}(a). The atom Fe1 with an up-arrow at the corner and Fe2 at the centre denotes the spin-up and spin-down magnetic moments. The spin density around the two magnetic atoms is represented by a transparent cloud like $d$-wave symmetry (denoted by cyan and red colours)[see Fig.\ref{fig:spin}(b,c)]. As we have observed, the volume of the spin density reduced on application of the compressive pressure, which may be related to a decrease in NRSS. The optimized lattice parameters were adopted for the calculations. The variations of the lattice parameter and unit cell volume versus applied pressure are shown in the supplementary material, Fig. S1. The computed value of lattice parameters and unit cell volume shows fairly good agreement with the experimental reports\cite{poffo2012structural}.

For further analysis of the dynamical stability, we have performed the phonon calculations. The phonon dispersion relationships and the projected density of states (DOS) of FeSb${_2}$ under external pressures ranging from 0.0 to 10 GPa reveal a pressure-dependent evolution of vibrational stability in this altermagnetic system. At ambient pressure (0.0 GPa), the phonon spectrum does not show imaginary frequencies in the Brillouin zone, confirming dynamical stability. The acoustic modes below 100 $cm^{-1}$ and the distinct optical branches up to 270 $cm^{-1}$ are well separated, with DOS projections indicating dominant Sb contributions in the lower frequency region, consistent with the mass and bonding asymmetries as in  Fig.\ref{fig:phon}(a). As pressure increases to 1.0 GPa, the entire phonon spectrum exhibits a moderate blue shift, reflecting a stiffening of the lattice and a compression-induced enhancement of interatomic force constants. The structure of the phonon band remains real-valued with no soft modes, indicating that FeSb${_2}$ retains its dynamical stability in this low-pressure regime. At 4.0 GPa, the phonon branches shift further upward in frequency, and the optical modes become slightly more dispersive, especially along the $\Gamma$-$A$ and $L$-$H$ directions, which can be seen in in Fig.\ref{fig:phon}(c). Despite the increasing pressure, the absence of imaginary phonon modes confirms that the material remains dynamically stable up to this pressure threshold. When pressure is further increased to 10 GPa, still no imaginary phonon modes were observed, a point indicating stiffening of the interatomic force constant and dynamical stability. The highest optical mode at $\Gamma$-point is around 280 $cm^{-1}$. The total and partial DOS of the phonon continues to show significant Sb character in low-frequency modes, with notable Fe1 and Fe2 contributions at intermediate frequencies, suggesting enhanced lattice participation of iron atoms under compression. The preservation of vibrational stability across this pressure range reinforces the mechanical robustness of FeSb${_2}$ and highlights the potential resilience of its crystalline phase against the external perturbations. Such phonon behaviour under pressure is in line with the theoretical predictions of narrow-gap correlated semiconductors that exhibit complex spin-lattice coupling\cite{ali2014large}.

\subsection{Electronic Properties}
Our previous work, which identified the ground state NRSS in FeSb$_2$\cite{phillips2025electronic}, has motivated us to extend this work to explore the unique electronic behaviour under applied pressures. The ground state antiferromagnetic coupling between Fe1 and Fe2 has already been confirmed by experimental observations, and the ground state energy comparison of different magnetic configurations from our DFT calculations is shown in Fig.S2. 

\begin{figure}\centering
	\includegraphics[width=1\linewidth]{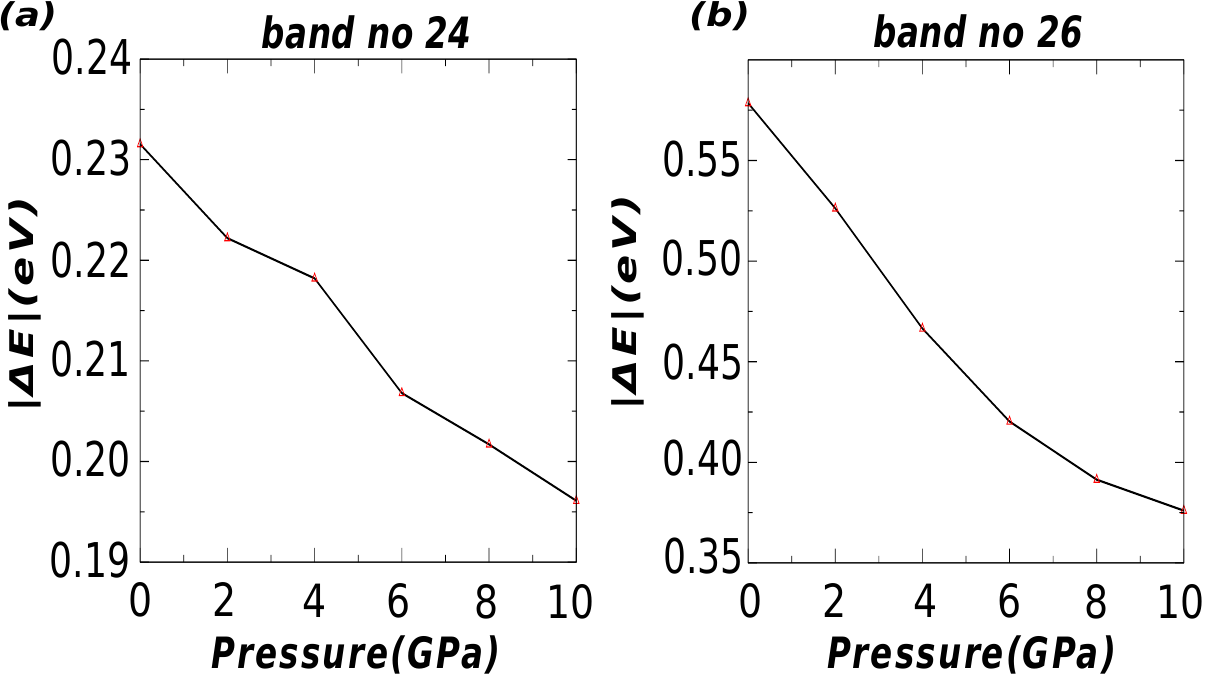}
	\caption{Variation of splitting energy under applied pressure in the range of 0 to 10 GPa for band indices 24 and  26.}
	\label{fig:NRSS}
\end{figure}

\begin{figure}
	\centering
	\includegraphics[width=1.0\linewidth]{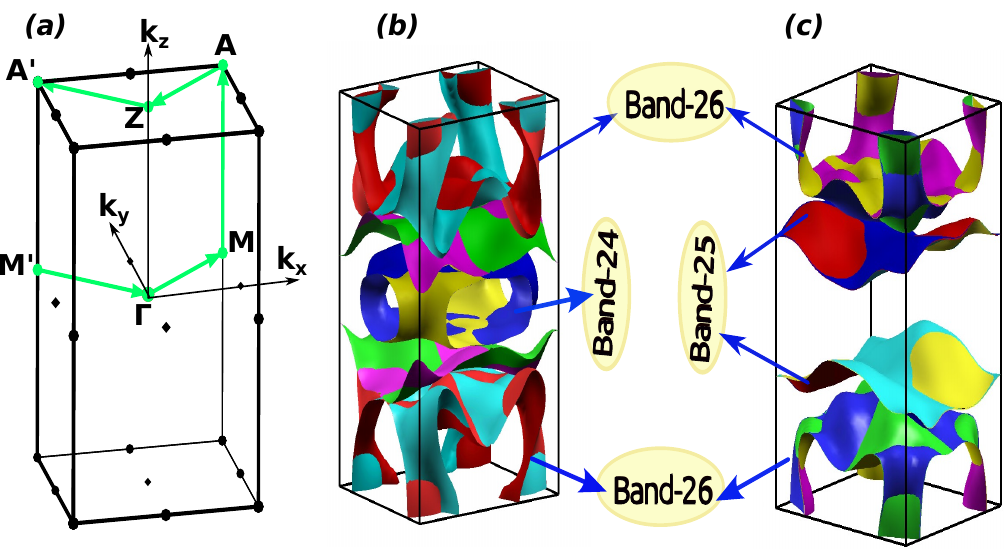}
	\caption{(a) First Brillouin-zone with high symmetry points M'-$\Gamma$-M-A-Z-A' and Fermi surface of FeSb$_2$ calculated at (b) 0.25 GPa and (c) 10.0 GPa (displaying two distinctive colours in each band representing spin-up and spin-down bands)}
	\label{fig:fermi}  
\end{figure}

To investigate the influence of external pressure on the electronic structure and the emergence of altermagnetism in FeSb${_2}$, we computed its spin-resolved band structures under various pressures (0.25--10.0 GPa), as shown in Fig.\ref{fig:band}. Under ambient conditions, FeSb${_2}$ exhibits motif pair anisotropy accompanied by collinear antiferromagnetic coupling between the two Fe atoms. The local environments surrounding the Fe${_1}$ and Fe${_2}$ atoms are related by a two-fold rotational symmetry around the [010] direction, followed by half a unit cell translation[$C_2||C_{2y}t$], while the $\mathcal{T}$- symmetry flips the spin quantization axis in the non-relativistic limit\cite{mazin2021prediction}. This combined action of spatial and $\mathcal{T}$- symmetries underpins the altermagnetic order observed in the system. The complete analysis of the electronic structure of FeSb${_2}$ at ambient conditions has already been performed using spin-polarised and SOC approaches and reported it as a potential altermagnet candidate\cite{phillips2025electronic}. The dispersion of the band was evaluated along the optimal path connecting high symmetry points within the first Brillouin zone, specifically M'-$\Gamma$-M-A-Z-A', as illustrated in Fig.\ref{fig:fermi}(a).  This path was selected to capture the key features of the electronic profile under various symmetry operations. The spin-polarised band structure reveals alternating features between the spin-up and spin-down channels, despite the collinear antiferromagnetic (AFM) coupling between the two Fe atoms and net zero magnetization. Interestingly, the dispersion does not resemble that of a typical ferromagnetic band structure, which exhibits constant splitting across the Brillouin zone due to exchange coupling or relativistic spin–orbit coupling (SOC)\cite{song2025altermagnets}. The presence of non-relativistic spin splitting (NRSS) near the high-symmetry points in the fully compensated magnet in the absence of SOC constitutes a hallmark of AM ordering. The strength of spin-splitting ($|\Delta E|$) around the Fermi energy is determined by subtracting the energy for the up and down spins, that is, E($\uparrow$) - E($\downarrow$) = $|\Delta E|$.\cite{vsmejkal2022beyond, vsmejkal2022emerging, vsmejkal2020crystal, yuan2023degeneracy, hayami2019momentum}

From the spin-resolved electronic band structures we have observed the crossing of the Fermi level by band-24, band-25, and band-26, which contribute to the Fermi surface topology at 0.25 GPa [see Fig.\ref{fig:band}(a)]. These bands exhibit the prominent spin splitting around the Fermi level with a clear separation between spin-up and spin-down states, supporting altermagnetic behaviour. The nondegenerate band at the $\Gamma$ and A-point forms a node due to the crossover of spin-up and spin-down bands. However, on either side of $\Gamma$ and A-point there lies a distinctive spin-splitting between the spin-up and spin-down components along M'-$\Gamma$-M and A-Z-A' high-symmetry paths [see Fig.\ref{fig:band}(a)]. Such a spin-splitting around the Fermi level indicates an anisotropic spin polarisation, crucial for spintronics. Band 26 exhibits the largest splitting along the A-Z symmetry, followed by band 24 along $\Gamma$-M, while band 25 shows moderate splitting above the Fermi energy. The NRSS along $\Gamma$-M (band-24) and A-Z (band-26) symmetries are $|\Delta E|$=0.23 eV and $|\Delta E|$=0.57 eV, respectively. From the analysis of the orbital-projected density of states (pdos) we have noticed that these bands are predominantly derived from Fe-3\textit{d} orbitals, particularly d$_{yz}$ orbital [see Supplementary Figure S3]. 

On increasing the pressure up to 1.0~GPa and 4.0~GPa, the bands 24 and 26 continue to preserve the spin splitting with decreasing values of $|\Delta E|$. At 1~GPa, both bands 24 and 26 show the distinct spin-up and spin-down components near the Fermi level, along $\Gamma$--$M$ and A-Z directions, respectively. However, the $|\Delta E|$ is slightly reduced as compared to that at 0.25 GPa. At the same time, we have observed the shifting of the node (formed by the crossover of spin up and spin down bands) towards the lower energy by 0.1 eV at $\Gamma$ and Z-point at 1.0 GPa [see Fig.\ref{fig:band}(b)]. This trend becomes more prominent at 4~GPa, where the spin separation further decreases and the bands begin to overlap more, indicating a suppression of altermagnetic effects under pressure [see Fig.\ref{fig:band}(a--c)]. On further increasing the pressure up to 10.0 GPa, the $|\Delta E|$ value drops from  $\sim$0.23 eV to $\sim$0.18 eV [see Fig.~\ref{fig:NRSS}(a)]. Notably, the reduction is more significant for band 26, where $|\Delta E|$ drops from approximately 0.57~eV to 0.37~eV at the same pressure range (0.25--10 GPa) [see Fig.~\ref{fig:NRSS}(b)], indicative of stronger sensitivity to external pressure as compared to band 24. The diminishing $d$-orbital character at the Fermi level with the applied pressure can be attributed to the suppression of the spin splitting, consistent with the previous reports.\cite{fan2025high, devaraj2024interplay,tamang2025first} The node (band-24) at $\Gamma$-point shift below the Fermi energy showing no Fermi-crossing at 10.0 GPa, whereas the bands 25 and 26 still cross the Fermi energy [Fig.~\ref{fig:band}(d)]. On the other hand, the $|\Delta E|$ value has been significantly suppressed and nearly degenerate, especially along the $\Gamma$--$M$ directions. This result shows the role of applied pressure in reducing the exchange splitting in FeSb$_2$ and diluting the essence of altermagnetism. The variations in band structures across the entire investigated pressure range are shown in Supplementary Fig.S4. 


Fig.\protect\ref{fig:fermi}, illustrates the Fermi surface of FeSb$_2$ at 0.25~GPa and 10.0~GPa, highlighting bands 24, 25, and 26 with spin-resolved pockets. At 0.25~GPa, band-24 forms a distinct yellow-colored pocket near the zone center, primarily derived from Fe-\textit{d} orbitals, which disappears entirely at 10~GPa, indicating a pressure-induced band shift below the Fermi level and leading to anisotropic electron pockets. Band-25, with contributions from both Fe-\textit{d} and Sb-\textit{p} orbitals, shows blue and cyan-colored lobes that evolve into more coherent structures under pressure, suggesting enhanced hybridization. Band-26, characterized by complex multi-colored features (red, cyan, purple), remains present across both pressures with significant spin splitting, reflecting robust spin polarization. The observed Fermi surface evolution underlines the sensitivity of orbital contributions to external pressure and the potential tunability of altermagnetic features in FeSb\textsubscript{2}.

\begin{figure*}
	\centering
	\includegraphics[width=0.46\linewidth]{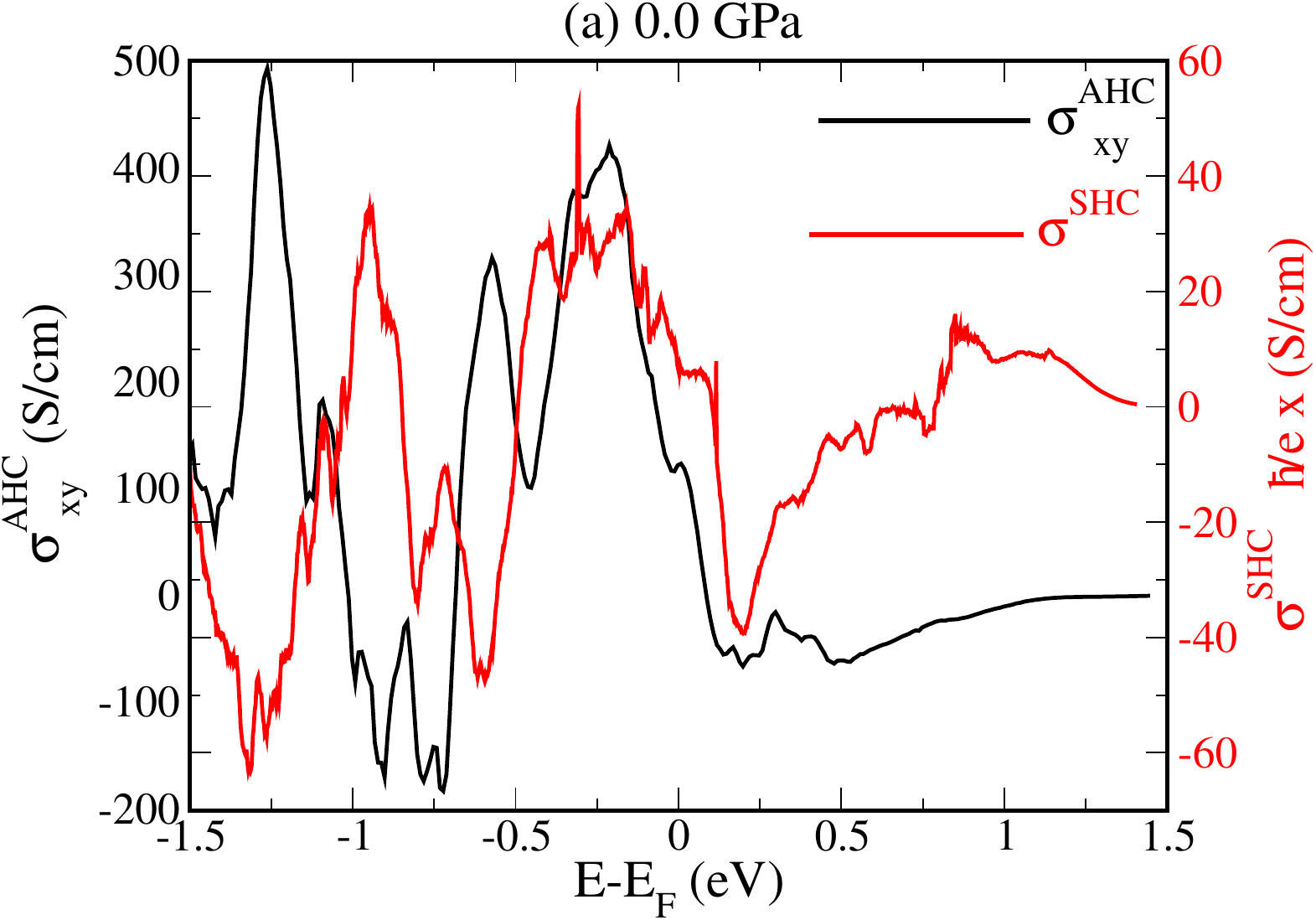}
	\includegraphics[width=0.46\linewidth]{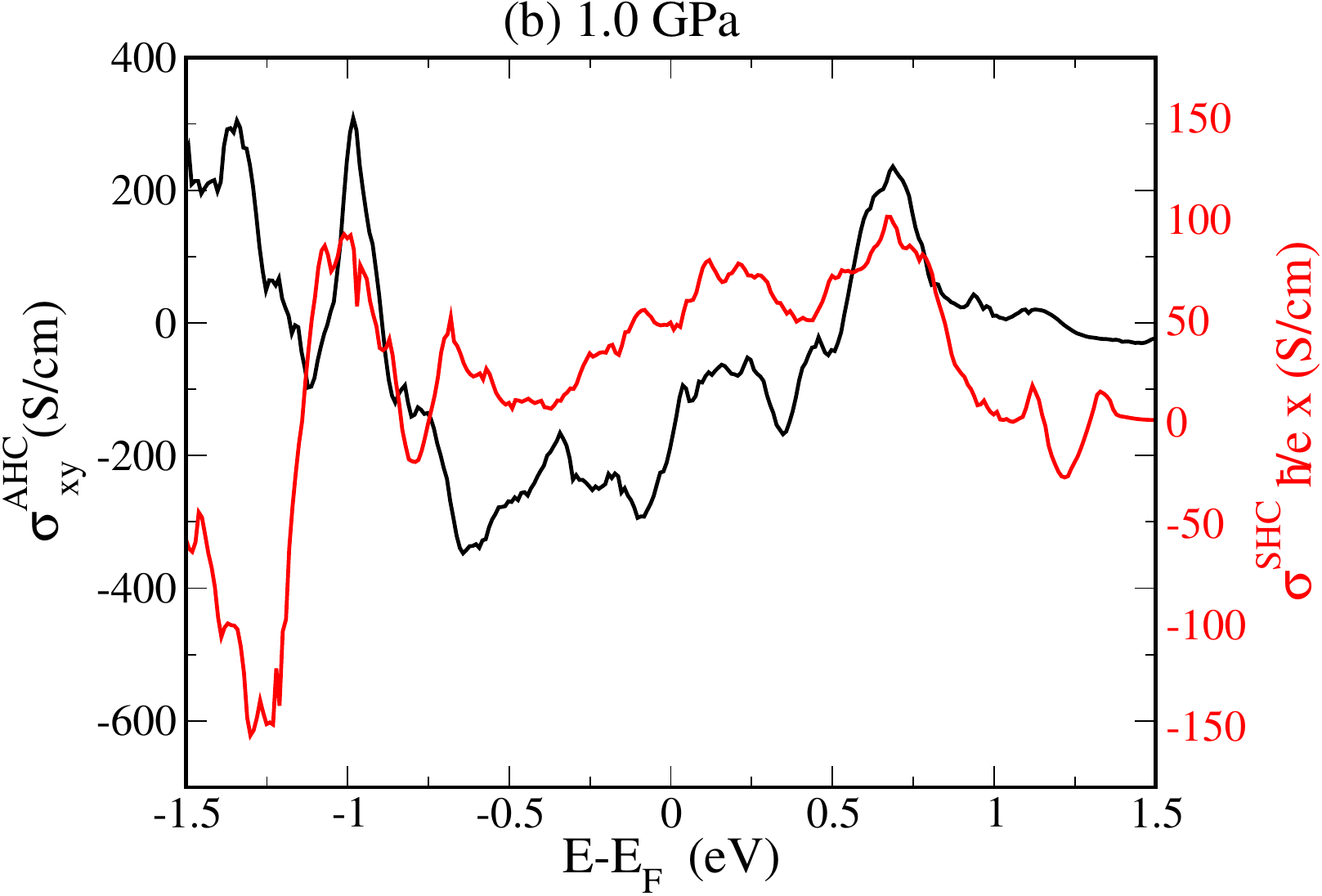}
	\includegraphics[width=0.46\linewidth]{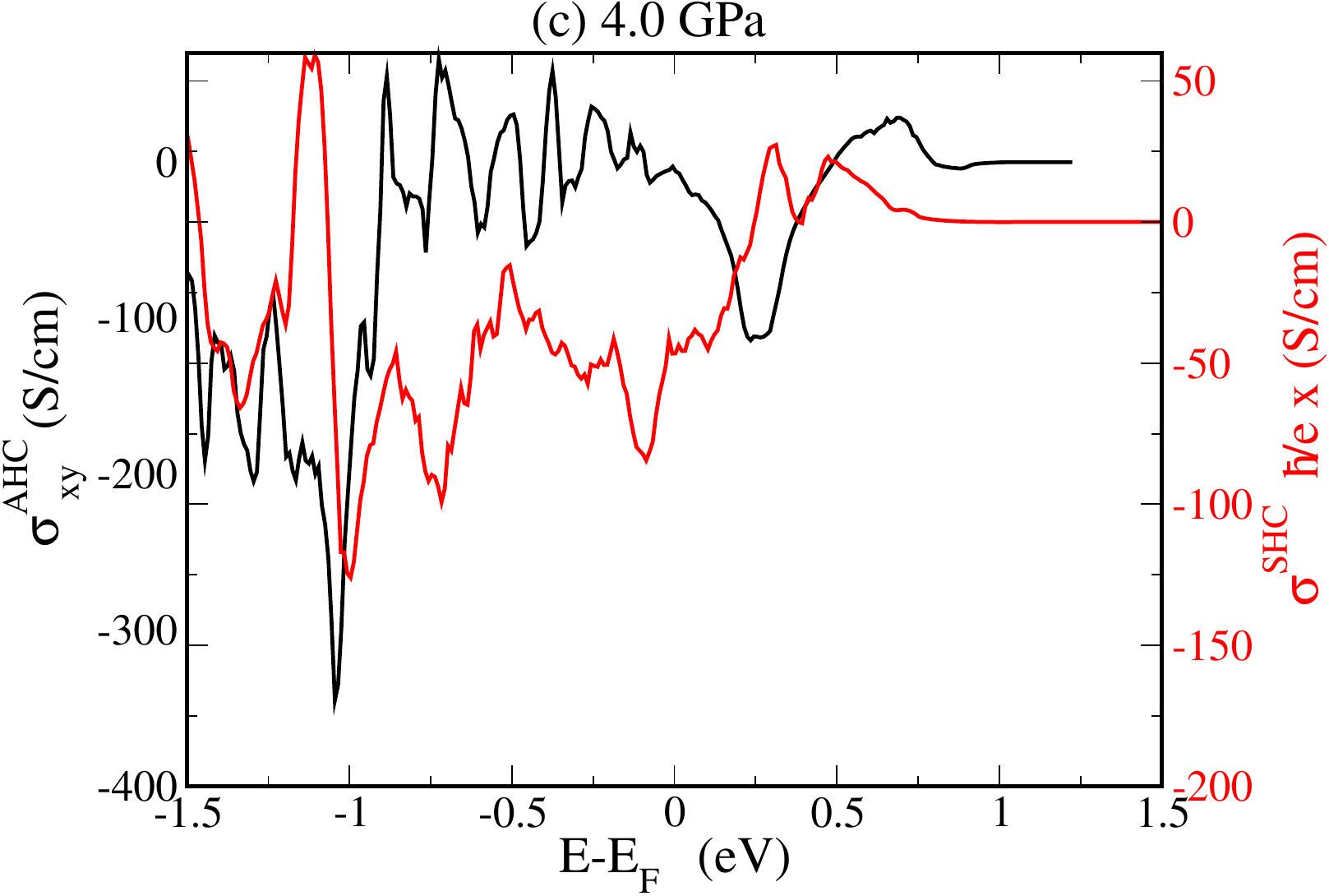}
	\includegraphics[width=0.46\linewidth]{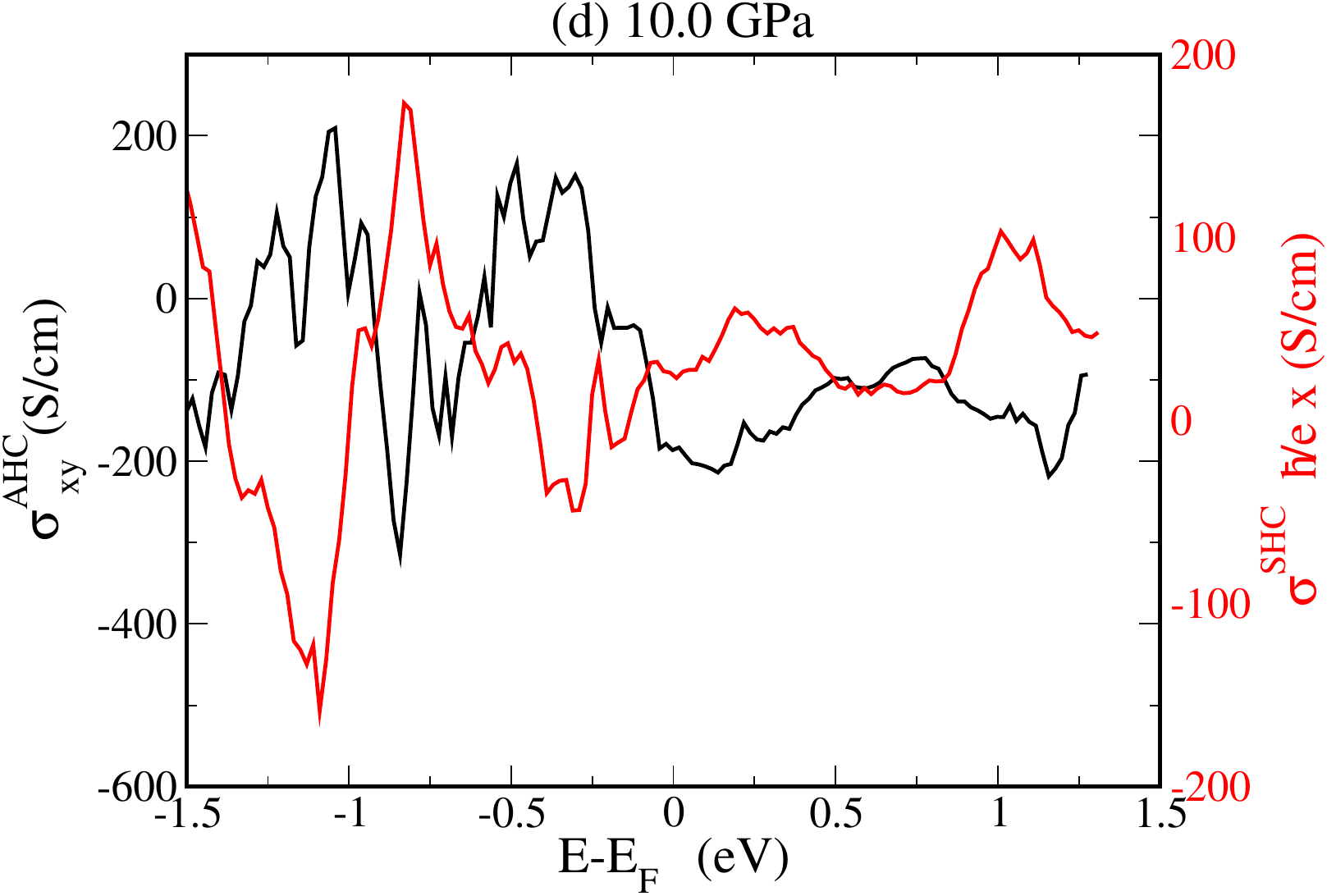}
	\caption{Anomalous Hall conductivity (Scaling on the left side) and Spin Hall conductivity (Scaling on the right side) of FeSb${_2}$ calculated at different pressures: (a) 0 GPa, (b) 1 GPa, (c) 4 GPa, and (d) 10 GPa. The AHC, $\sigma_{xy}$, is plotted as a function of energy relative to the Fermi level (E-E${_F}$). At ambient pressure, sharp peaks reflect strong Berry curvature near the Fermi level, consistent with altermagnetic band topology. With increasing pressure, the spectral features evolve significantly, showing redistribution, suppression,  and broadening of AHC signatures, indicative of pressure-tuned topological transitions in the electronic structure.}
	\label{fig:ahc}  
\end{figure*}

\subsection { Anomalous and Spin Hall Effect}
The anomalous Hall effect (AHE) serves as a macroscopic probe of $\mathcal{T}$-symmetry breaking in magnetic systems. Traditionally associated with ferromagnets and non-collinear antiferromagnets, AHE has recently been observed in altermagnets. These systems exhibit alternating spin polarization in momentum space, enabling $\mathcal{T}$-symmetry to break without net magnetization. Unlike ferromagnets, where AHE is allowed in symmetry in all orientations, altermagnets display AHE only for specific N\'eel vector directions relative to the crystal axes. In $\alpha$-MnTe, for instance, AHE emerges when the N\'eel vector aligns with $\langle 1\bar{1}00 \rangle$ axes, showing hysteresis and remanent signal\cite{gonzalez2023spontaneous}. In contrast, in tetragonal RuO$_2$, AHE is forbidden by symmetry when the N\'eel vector lies along the [001] axis, being allowed by symmetry only when the vector has in-plane components. \cite{tschirner2023saturation} Consequently, no AHE is observed in out-of-plane field configurations, even under spin-flop conditions, as the overall symmetry remains preserved. Likewise, spin Hall effect  (SHE) plays a key role in spin-to-charge conversion that converts a charge current into spin current\cite{sinova2015spin}. The spin Hall conductivity (SHC) is a function of the Fermi level, so the electrons around the E$_F$ only contribute to spin transport activities.  In FeSb${_2}$, SHC is driven by spin-orbit entanglement of the Fe 3\textit{d} and Sb 5\textit{p} states, especially where spin-up and spin-down components are spatially separated due to SOC. The AHC  describes the generation of a transverse dissipationless current in response to a longitudinal electric field, arising intrinsically from the Berry curvature of occupied electronic bands rather than from external magnetic fields\cite{devaraj2024interplay,vsmejkal2020crystal}. In FeSb${_2}$, we present the AHC spectra for the orientation of the magnetization [0 0 1] direction under applied pressures of 0.25-10 GPa, computed via the Berry curvature formalism with included SOC (Fig.\ref{fig:ahc}).

At ambient pressure (0 GPa), a pronounced positive peak is observed just below the E$_F$, reaching values above \(+400~\mathrm{S/cm}\) followed by rapid oscillations and decay and the AHC at the E$_F$ is found to be approximately $130~S\text{cm}^{-1}$, which is in good agreement with the previous reported  value\cite{mazin2021prediction}.  The  SHC at the E$_F$ is about $ +220 ~S\text{cm}^{-1}$, indicating strong Berry curvature contributions from altermagnetic band crossings. These features correspond to sharp Berry curvature contributions from nodal structures near the E$_F$, consistent with our prior analysis of altermagnetic band crossings, as shown in Fig.\ref{fig:band}.  Fig. \ref{fig:ahc}(b) shows the result of the AHC  and SHC at 1 GPa, showing broader and more asymmetric features. The peak near \(-1~\mathrm{eV}\) diminishes in height, and both conductivities change sign and magnitude, with the AHC dropping to around \(-105~\mathrm{S/cm}\) whereas SHC reduces to approximately   \(+50~\mathrm{S/cm}\) at E$_F$. This suggests enhanced Berry curvature contributions from deeper bands, indicating pressure-induced band inversions.

Increasing the pressure further to 4 GPa,  the overall AHC is markedly suppressed, with maximum values not exceeding \(\pm 250~\mathrm{S/cm}\). The spectral weight shifts further into the negative-energy region, and the structure becomes more diffuse. The AHC becomes more negative, reaching \(\- 250~\mathrm{S/cm}\) and the SHC further diminishes to near \(0\,\mathrm{S/cm}\)  at E$_F$, suggesting a pressure-induced suppression of spin-orbit-driven transport. This indicates a transition regime where the dominant topological features are modified or suppressed by pressure. At 10 GPa, the AHC spectrum becomes highly oscillatory with strong negative and positive contributions between $-0.75~\mathrm{eV}$ and $0~\mathrm{eV}$ and the AHC at the E$_F$ is found to be approximately \(-150~\mathrm{S/cm}\) while the SHC slightly recovers to roughly \(60~\mathrm{S/cm}\), suggesting that multiple band crossings and their associated Berry curvatures are active.  Despite the spectral complexity, significant AHC persists, implying robust topological character even at high compression, as shown in Fig.\ref{fig:ahc}(d). The coexistence and tunability of both AHC and SHC highlight the dual spin-charge transport nature of this material and its potential for pressure-controlled spintronic functionalities.


\section{Summary}
In this study, we have systematically examined the influence of  pressure on the electronic, magnetic, and transport properties of FeSb${_2}$, a recently confirmed altermagnetic material. Our first-principles calculations show that the system remains dynamically stable up to 10 GPa, as characterised by phonon dispersion analysis. With increasing pressure, FeSb${_2}$ undergoes a transition to a semimetallic regime and significant restructuring of the electronic bands near the Fermi level. This electronic evolution is accompanied by an enhancement in staggered orbital magnetisation, 
reflecting the robustness of the altermagnetic state under compression. Most notably, the anomalous Hall conductivity (AHC) exhibits a strong pressure dependence, evolving from a sharply peaked positive response at ambient conditions to a spectrally broadened and negative response at high pressure. This behaviour is attributed to pressure-induced band crossings and the associated redistribution of Berry curvature. Our results highlight the potential of FeSb${_2}$  as a model system for exploring pressure-tunable altermagnetism and Berry phase-driven transport,  providing a pathway toward functional spintronic materials with controllable topological properties.

\begin{acknowledgments}
	\textbf{DPR} acknowledges the Science \& Engineering Research Board (SERB), New Delhi Govt. of India via File Number: SIR/2022/001150. \\
	\textbf{RT} acknowledges the University Grants Commission (UGC), India, for the Junior Research Fellowship (JRF), ID No. 231620066332. \\
	\textbf{K. S.} was supported in part by the U.S. Department of Energy, Office of Science, Office of Workforce Development for Teachers and Scientists (WDTS) under the Visiting Faculty Program (VFP) at Los Alamos National Laboratory, administered by the Oak Ridge Institute for Science and Education. The work at West Texas A$\&$M University (WTAMU) is supported by the Welch Foundation (Grant No. AE-0025) and the National Science Foundation (Award No. 2336011). The computations were performed on the WTAMU HPC cluster, which was funded by the National Science Foundation (NSF CC* GROWTH 2018841).
\end{acknowledgments}

\section*{Author contributions}
\textbf{Shalika R. Bhandari:} Formal analysis, Visualization, Validation, Literature review, Writing-original draft, writing-review \& editing.\\
\textbf{Keshav Shrestha:} Formal analysis, Visualization, Validation, Literature review,  writing-review \& editing.\\
\textbf{R. Tamang:} Formal analysis, Visualization, Validation, Literature review, writing-review \& editing.\\
\textbf{Samy Brahimi}:Formal analysis, Visualization, Validation, writing-review \& editing.\\
\textbf{Samir Lounis}: Supervision, Resources, Formal analysis, Visualization, Validation, writing-review \& editing.
\textbf{D. P. Rai:} Formal analysis, Visualization, Validation, Writing-original draft, writing-review \& editing. \\

\section*{Data Availability Statement}
Data available within the article or its supplementary materials	

\nocite{*}
\bibliography{rsc}

\end{document}